\documentclass[conference, a4paper]{IEEEtran}
\IEEEoverridecommandlockouts
\usepackage{cite}
\usepackage{amsmath,amssymb,amsfonts,mathtools}
\usepackage{algorithmic}
\usepackage{graphicx}
\usepackage{textcomp}
\usepackage{svg}
\usepackage{multirow}
\usepackage{pifont}
\usepackage{xcolor}
\usepackage{placeins}
\usepackage{enumitem}
\usepackage{siunitx}
\usepackage{listings}
\DeclareFixedFont{\ttb}{T1}{txtt}{bx}{n}{10} 
\DeclareFixedFont{\ttm}{T1}{txtt}{m}{n}{10}  
\usepackage{color}
\definecolor{deepblue}{rgb}{0,0,0.5}
\definecolor{deepred}{rgb}{0.6,0,0}
\definecolor{deepgreen}{rgb}{0,0.5,0}
\newcommand\pythonstyle{\lstset{
language=Python,
basicstyle=\ttm,
morekeywords={vGate,iTargetSet,vGateRead,vGateReset,iTarget,vRead,iTargetReset},
keywordstyle=\ttb\color{deepblue},
emph={formCell,resetCell,setCell,readCell},          
emphstyle=\ttb\color{deepred},    
stringstyle=\color{deepgreen},
frame=tb,                         
showstringspaces=false
}}
\lstnewenvironment{python}[1][]
{
\pythonstyle
\lstset{#1}
}
{}

\usepackage{stfloats} 
\usepackage{subcaption}
\usepackage{tikz}
\usepackage{xspace}
\usepackage{circledsteps}
\usepackage{chemmacros}

\usepackage{tikz}
\usepackage[]{pgfplots}
\usepackage{pgfplotstable}
\pgfplotsset{compat=1.18}
\pgfplotsset{
    General/.style={
        font=\footnotesize,
        width=\linewidth,
        height=4cm,
        xtick pos=left,
        xtick align=outside,
        ytick pos=left,
        ytick align=outside,
        ymajorgrids=true,
        grid style=dashed,
        legend cell align={left},
        style=thin,
    },
    BarConfig/.style={
        General,
        ymin=0,
        enlarge x limits=0.25,
        xtick=data,
        xticklabel style={rotate=30, anchor=east},
        xlabel shift=-6pt,
        bar width=8pt,
    },
    ScatterConfig/.style={
      General,
      only marks,
      xtick=data,
      xticklabel style={rotate=90, anchor=east},
    }
}

\usepgfplotslibrary{statistics}
\makeatletter
\pgfplotsset{
    boxplot prepared from table/.code={
        \def\tikz@plot@handler{\pgfplotsplothandlerboxplotprepared}%
        \pgfplotsset{
            /pgfplots/boxplot prepared from table/.cd,
            #1,
        }
    },
    /pgfplots/boxplot prepared from table/.cd,
        table/.code={\pgfplotstablecopy{#1}\to\boxplot@datatable},
        row/.initial=0,
        make style readable from table/.style={
            #1/.code={
                \pgfplotstablegetelem{\pgfkeysvalueof{/pgfplots/boxplot prepared from table/row}}{##1}\of\boxplot@datatable
                \pgfplotsset{boxplot/#1/.expand once={\pgfplotsretval}}
            }
        },
        make style readable from table=lower whisker,
        make style readable from table=upper whisker,
        make style readable from table=lower quartile,
        make style readable from table=upper quartile,
        make style readable from table=median,
        make style readable from table=average,
        make style readable from table=lower notch,
        make style readable from table=upper notch,
        make style readable from table=draw position,
}
\makeatother

\usepackage[ruled, linesnumbered]{algorithm2e}
\SetKwComment{Comment}{//}{}

\usepackage[hyperfootnotes=false, colorlinks=false, pdfborder={0 0 0}]{hyperref}
\usepackage[capitalise]{cleveref}
\usepackage[printonlyused,nolist]{acronym}
\def\BibTeX{{\rm B\kern-.05em{\sc i\kern-.025em b}\kern-.08em
    T\kern-.1667em\lower.7ex\hbox{E}\kern-.125emX}}

\usepackage[all]{background}
\usepackage{stackengine}
\setstackEOL{\\}
\setstackgap{L}{\normalbaselineskip}
\SetBgContents{\color{gray}{\tiny \Longstack{PREPRINT - Accepted by 25th IEEE Latin American Test Symposium (LATS) 2024, Maceio, Brazil}}}
\SetBgPosition{3.5cm,1cm}
\SetBgOpacity{1.0}
\SetBgAngle{0}
\SetBgScale{1.8}

\begin{document}

\bstctlcite{IEEEexample:BSTcontrol}

\definecolor{c0}{HTML}{1b85b8}
\definecolor{c1}{HTML}{5a5255}
\definecolor{c2}{HTML}{559e83}
\definecolor{c3}{HTML}{ae5a41}
\definecolor{c4}{HTML}{c3cb71}
\definecolor{c5}{HTML}{bfb5b2}

\definecolor{cweighted}{HTML}{a72090}
\definecolor{cminerror}{HTML}{008080}

\begin{acronym}
    \acro{rram}[ReRAM]{Resistive Random Access Memory}
    \acro{cim}[CIM]{Computing-in-Memory}
    \acro{mvm}[MVM]{Matrix-Vector Multiplication}
    \acro{nbb}[NBB]{NeuroBreakoutBoard}
    \acro{nvm}[NVM]{Non-Volatile Memories}
    \acro{tia}[TIA]{TransImpedance Amplifier}
    \acro{hrs}[HRS]{High Resistive State}
    \acro{lrs}[LRS]{Low Resistive State}
    \acro{wl}[WL]{Word Line}
    \acro{bl}[BL]{Bit Line}
    \acro{nn}[NN]{Neural Network}
    \acro{adc}[ADC]{Analog-to-Digital Converter}
    \acro{dac}[DAC]{Digital-to-Analog Converter}
    \acro{api}[API]{Application Programming Interface}
    \acro{ispva}[ISPVA]{Incremental Step Pulse Program and Verify Algorithm}
    \acro{nmos}[NMOS]{Negative channel Metal-Oxide Semiconductor}
    \acro{mos}[MOS]{Metal-Oxide Semiconductor}
    \acro{cmos}[CMOS]{Complementary Metal-Oxide Semiconductor}
    \acro{mim}[MIM]{Metal-Insulator-Metal}
    \acro{oxram}[OxRAM]{Oxide-based Random Access Memory}
    \acro{sl}[SL]{Source Line}
    \acro{c2c}[C2C]{Cycle-to-Cycle}
    \acro{d2d}[D2D]{Device-to-Device}
    \acro{cdf}[CDF]{Cumulative Distribution Function}
    \acro{cf}[CF]{Conductive Filament}
    \acro{wm}[WM]{Window Margin}
    \acro{if}[IF]{Incremental Form}
    \acro{ifv}[IFV]{Incremental Form and Verify}
    \acro{vcm}[VCM]{Valence Change Memory}
    \acro{ecm}[ECM]{Electrochemical Mechanism}
    \acro{tcm}[TCM]{Thermochemical Mechanism}
    \acro{uart}[UART]{Universal Asynchronous Receiver/Transmitter}
    \acro{protobuf}[Protobuf]{Protocol Buffer}
    \acro{fmc}[FMC]{FPGA Mezzanine Card}
\end{acronym}

\def\finalpaper{1} 
\title{A Fully Automated Platform for Evaluating ReRAM Crossbars}
\if\finalpaper1
    \author{
        \IEEEauthorblockN{Rebecca Pelke\IEEEauthorrefmark{1}, Felix Staudigl\IEEEauthorrefmark{1},
        Niklas Thomas\IEEEauthorrefmark{1}, Nils Bosbach\IEEEauthorrefmark{1},
        Mohammed Hossein\IEEEauthorrefmark{1}, Jose Cubero-Cascante\IEEEauthorrefmark{1},
        \\Leticia Bolzani Poehls\IEEEauthorrefmark{2},
        Rainer Leupers\IEEEauthorrefmark{1}, and Jan Moritz Joseph\IEEEauthorrefmark{1}}
        \IEEEauthorblockA{
            \IEEEauthorrefmark{1}\textit{Institute for Communication Technologies and Embedded Systems, RWTH Aachen University}\\
            \IEEEauthorrefmark{2}\textit{Chair of Integrated Digital Systems and Circuit Design, RWTH Aachen University}\\
            \{pelke, staudigl, thomas, bosbach, hossein, cubero, leupers, joseph\}@ice.rwth-aachen.de\\
            poehls@ids.rwth-aachen.de}
            \vspace{-1cm}
            \thanks{This work was partially funded by the Federal Ministry of Education and Research (BMBF, Germany) in the project NeuroSys (Project Nos. 03ZU1106CA) and NEUROTEC (project Nos. 16ES1134 and 16ES1133K).}
    }
\else
    \author{
      \IEEEauthorblockN{Authors are removed for submission version}
      \\
      \\
      \IEEEauthorblockA{Affiliations are removed for submission version}
      \vspace{0cm}
      \\
    }
\fi
\IEEEoverridecommandlockouts
\IEEEpubid{\makebox[\columnwidth]{979-8-3503-6555-9/24/\$31.00 \copyright2024 European Union\hfill}
\hspace{\columnsep}\makebox[\columnwidth]{ }}

\maketitle

\begin{abstract}
\Ac{rram} is a promising candidate for implementing \ac{cim} architectures and neuromorphic circuits.
\ac{rram} cells exhibit significant variability across different memristive devices and cycles, necessitating further improvements in the areas of devices, algorithms, and applications.
To achieve this, understanding the stochastic behavior of the different \ac{rram} technologies is essential.
The \ac{nbb} is a versatile instrumentation platform to characterize \acl{nvm} (NVMs).
However, the \ac{nbb} itself does not provide any functionality in the form of software or a controller.
In this paper, we present a control board for the \ac{nbb} able to perform reliability assessments of 1T1R \ac{rram} crossbars.
In more detail, an interface that allows a host PC to communicate with the \ac{nbb} via the new control board is implemented.
In a case study, we analyze the \ac{c2c} variation and read disturb TiN/Ti/HfO\textsubscript{2}/TiN cells for different read voltages
to gain an understanding of their operational behavior.

\end{abstract}
\vspace{-0.25cm}

\acresetall 

\section{Introduction}
There is an ever-growing demand for efficient \ac{nn} inference.
Due to the computationally and data-intensive workloads,
traditional computers quickly encounter the
von Neumann bottleneck~\cite{zou2021breaking} and the memory-wall~\cite{huang2020memory}.
One possible solution for addressing this problem is the adoption of \ac{cim} architectures. 
The concept involves statically storing the weights of \acp{nn}
in non-volatile \ac{cim} cells.
The computations take place within the memory,
tackling the challenges of traditional computing architectures~\cite{wei2022trends, baroni2022energy}.
\ac{rram} is a promising candidate for \ac{cim} due to its high device density,
scalability, switching performance, low power consumption,
and \ac{cmos} manufactuirng process compatibility~\cite{hazra2021optimization, meena2014overview, vetter2015opportunities}.

The downside of \ac{rram} is poor uniformity between cycles and devices,
known as \ac{c2c} and \ac{d2d} variabilities~\cite{6724674}.
\ac{c2c} variation is associated with the stability and reliability of \ac{rram} devices over repeated cycles of use.
It is an intrinsic property of the technology~\cite{wong2012metal,yu2011stochastic}.
\ac{d2d} variation is associated with manufacturing and production consistency.
Manufacturing deviations in thickness, length, width, and surface roughness of the \ac{rram} cells increase the \ac{d2d} variation~\cite{brum2021evaluating}.
Note that these problems prevent the rapid commercialization and adoption of \ac{rram} for implementing emerging applications.
To advance \ac{rram} technology, there are several topics to explore, including
\begin{enumerate}
    \item enhancing \ac{rram} devices~\cite{fang2018improvement, wu2018methodology, kempen202150x}.
    \item improving program and read algorithms~\cite{liu2014uniformity, baroni2021tackling, bengel2022reliability}.
    \item increasing the fault tolerance of applications~\cite{xia2017fault, xia2017stuck, chen2017accelerator}.
\end{enumerate}

Regardless of the topic,
understanding the characteristics of individual devices is essential for properly implementing and optimizing these emerging applications.
Thus, automated measurement processes are required to reduce the need for manual and time-consuming laboratory measurements.
Different systems have been developed for this purpose~\cite{berdan2015mu, cayo2021design, de2020development,staudigl2023neuroboard}, such as the \ac{nbb}.
The \ac{nbb} can measure a wider range of devices due to its flexibility~\cite{staudigl2023neuroboard}.

The flexibility is achieved through a combination of the interconnect matrix, custom-designed \acp{tia},
and the omission of already-fixed control mechanisms on the \ac{nbb}.
This means that the components of the \ac{nbb} need to be controlled by an external device
because the \ac{nbb} itself is not yet a fully functional measurement system.
We aim to close this gap by developing a new platform able to perform reliability assessments of 1T1R \ac{rram} crossbars.
In this context, the main contributions of this paper are:

\begin{itemize}
    \item Development of a controller board that allows controlling the \ac{nbb} from a host PC, eliminating the need for supervision during measurements (increase of measurement consistency and time savings in the laboratory).
    \item Implementation of a new interface accessible from the host PC for easily configuring the target experiments.
    \item Validation of the proposed controller board and interface based on a case study aiming the analysis of read voltage impact on \ac{c2c} variability and read disturb of TiN/Ti/HfO\textsubscript{2}/TiN cells.
\end{itemize}

The paper is organized as follows:
\Cref*{sec:background} provides the required background related to \ac{rram}s' behavior as well as summarizes the main metrics adopted for evaluating \ac{rram} cells.
\cref*{sec:implementation} describes the proposed controller board and interface.
\Cref*{sec:casestudy} summarizes the results related to the performed case study's reliability assessment.
Finally, \cref*{sec:conclusion} draws a conclusion.

\section{Structure}
\label{sec:background}
This Section presents the background related to memristive devices.
It further introduces their main evaluation metrics.

\subsection*{Memristive Devices: Structure and Functionality}
\label{sec:rrambackground}
A memristive device is composed of a transition-metal-oxide layer, e.g., HfO\textsubscript{2}, ZrO\textsubscript{2}, Ta\textsubscript{2}O\textsubscript{5}, or TiO\textsubscript{2},
between two conducting electrodes~\cite{hazra2021optimization}.
After manufacturing, especially in oxide-based filamentary-type devices, memristive devices usually have very high electrical resistance.
A large voltage is required for the very first \textit{set} operation, also known as the \textit{forming} process~\cite{8714728}.
This process, a controlled soft breakdown, drastically reduces the device resistance enabling the resistance-switching behavior in the subsequent cycles for the filamentary regime~\cite{poehls2021review}.
The forming process directly impacts the lifespan, forming yield,
and \ac{c2c} variability of the memristor.
After forming, \textit{set} and \textit{reset} operations can be applied, which bring the device to the \ac{lrs} and \ac{hrs}, respectively~\cite{puglisi2014study}.

The authors of~\cite{grossi2016electrical} compared three forming algorithms:
Single pulse, \ac{if}, and \ac{ifv}.
The first method applies a single voltage pulse for forming memristors,
the second one applies multiple voltage pulses with increasing magnitudes,
and the third, \ac{ifv},
involves increasing voltage pulses combined with verification steps to monitor
the state of the cell and to prevent current overshoots.
They have found out that \ac{ifv} outperforms the other techniques in terms of
cell yield and switching behavior due to the tight control of the conductive filament creation.

In this work, the \ac{ispva}~\cite{perez2019toward}, an \ac{ifv} algorithm, is used.
The algorithm, illustrated in \cref*{fig:ispva}, applies voltage pulses with increasing magnitude to the electrodes
to change the resistive state.
After each programming pulse $\mathrm{V_{prog}}$,
a verification voltage $\mathrm{V_{verify}}$ is applied to measure if the desired current has been reached.
The algorithm contains adjustable parameters, e.g.,
the pulse duration $\mathrm{T_{pulse}}$, the start voltage $\mathrm{V_{start}}$,
the voltage increase $\mathrm{V_{step}}$, and the stop voltage $\mathrm{V_{stop}}$.
During \textit{set}, positive voltages are applied to
the \ac{wl} and \SI{0}{\volt} to the \ac{bl}.
During \textit{reset}, positive voltages are applied to
the \ac{bl} and \SI{0}{\volt} to the \ac{wl}.

\begin{figure}[b!]
    \centering
    \vspace{-0.5cm}
    \includesvg[inkscapelatex=false, width=.8\linewidth, keepaspectratio]{figures/ispva.svg}
    \caption{\ac{ispva}~\cite{perez2019toward} voltage sequence}
    \label{fig:ispva}
\end{figure}

\subsection*{\acp{rram}: Structure and Metrics}

In 1T1R crossbars, each \ac{rram} cell consists
of one transistor and one memristive device.
The transistor is used as a switch to disconnect the memristor from the
crossbar, which reduces sneak-path currents~\cite{mao2016optimizing}.
There are three types of 1T1R crossbars,
namely \textit{typical}, \textit{vertical},
and \textit{pseudo} crossbars~\cite{son2022study}.
They differ in the horizontal or vertical placement of the
\ac{sl}, \ac{wl}, and \ac{bl}.
The conductance value of the memristor $\mathrm{G_{ij}}$ represents the weight.
The read voltage applied to the \ac{wl} represents the input,
and the output current at the \ac{bl} corresponds to the result of the
\ac{mvm} between the weights and inputs.
The \ac{sl} is connected to the gate of the transistor.
\cref*{fig:crossbar} shows a pseudo crossbar, which is used later in the case study.
In pseudo crossbars, \ac{wl} and \ac{sl} are both placed in the horizontal direction.
\begin{figure}[t!]
    \centering
    \includesvg[inkscapelatex=false, width=.8\linewidth, keepaspectratio]{figures/crossbar.svg}
    \caption{1T1R crossbar (pseudo crossbar)}
    \label{fig:crossbar}
    \vspace{-0.4cm}
\end{figure}
%
%
The following properties are typically analyzed to characterize a \ac{rram} cell:
\begin{itemize}
    \item \textbf{\Acl{wm}}: $\mathrm{R_{OFF}/R_{ON}}$ ratio.
    \item \textbf{Retention}: Stability of the cell (at higher temperatures).
    \item \textbf{Endurance}: Maximum number of programming cycles $\mathrm{N_{cmax}}$ at a constant \ac{wm}.
    \item \textbf{Read disturb}: Resistance-drifts after multiple readouts.
    \item \textbf{Variation}: \ac{c2c} variation (different cycles) and \ac{d2d} variation (different devices).
\end{itemize}

Different physical compositions of the \ac{rram} cells result in
varying behaviors.
MO\textsubscript{x}/CuTe\textsubscript{x} shows a good retention,
while HfO\textsubscript{2}/CuTe\textsubscript{x} reports large \acp{wm} at a high endurance.
HfO\textsubscript{2}/Ti present a good trade-off between
endurance and retention~\cite{nail2016understanding}.
Related works have also revealed that programming algorithms
impact the switching behavior~\cite{perez2021variability},
variability, and reliability~\cite{baroni2022low, baroni2021tackling}.
Further, the read voltages have significant effects on read disturb
and \ac{c2c} variation.
The authors of~\cite{bengel2022reliability} analyzed this for Pt/ZrO\textsubscript{2}/Ta/Pt cells.
They have found out that it is advantageous to read in the \textit{reset} direction.
In the following study, we plan to examine this phenomenon for HfO\textsubscript{2}/Ti cells.
A particular focus is placed on the \ac{hrs} since a gradual \ac{hrs} drift
towards the \ac{lrs} is observed over cycling for this cell type~\cite{sassine2018optimizing}.
When $\mathrm{N_{cmax}}$ is exceeded, programming the \ac{hrs} becomes unfeasible,
i.e., the cell is broken.

\subsection*{The \acl{nbb} (\ac{nbb})}
\label{sec:nbb}
The aim of \ac{nbb} is to provide a customizable infrastructure for measuring \ac{nvm}~\cite{staudigl2023neuroboard}.
The \ac{nbb} includes a configurable interconnect matrix and circuitry for voltage pulse generation
and current measurement.
These components are explained in more detail in the following.

\subsubsection{Configurable Signal Interconnection Module}
The board employs arrays of multiplexers to distribute different signal lines or ground potential to each pin of the \ac{nvm} socket independently.
Around the socket, \num{68} \num{8}:\num{1} multiplexers are spread over the four sides of the socket (see \cref*{fig:overview}).
Each multiplexer receives \num{5} signals from the \ac{dac}, a ground line, and an external voltage line.
The south multiplexers are additionally connected to the current sensing module.

\subsubsection{Voltage Pulse Generator Module}
A \num{12}-\si{\bit}, \num{16}-\si{channel} \ac{dac} is used to generate the
different voltage levels needed for the crossbar operations.
The \num{16} channels can be used independently.
Programming \ac{rram} cells requires very short and precise voltage pulses~\cite{perez2019toward}.
For this purpose, additional analog switches with switching times below \SI{60}{\nano\second} are provided.

\subsubsection{Current Sensing Module}
Accurate current sensing is required for different tasks like reading the calculation result of a \ac{mvm}.
Furthermore, programming algorithms, such as \ac{ispva}, constantly require accurate feedback
about the current.
The currents can be measured at the \acp{bl} of the crossbar (see \cref*{fig:crossbar}).
For accurate sensing, the \ac{nbb} has a circuit called \ac{tia} sense,
which uses adjustable feedback resistors to translate currents into voltages.
The \ac{nbb} has \num{16} \ac{tia} sense circuits, one for each south pin.
To measure a wide range of currents, different feedback resistances from
\SI{43}{\ohm} to \SI{100}{\kilo\ohm} can be used.
\ac{tia} sense further includes an amplifier to adjust the voltage for the \ac{adc}.
The \ac{adc} has \num{8} \num{18}-\si{\bit} channels, which can be used simultaneously.
The input resistance is \SI{1}{\mega\ohm}.

The authors of~\cite{staudigl2023neuroboard} evaluated the measurement accuracy of the \ac{nbb}.
For measuring crossbar cells, we expect the resistances to range from \SI{1}{\kilo\ohm} to \SI{100}{\kilo\ohm}.
In this range, the measurement accuracy remains highly reliable,
with a relative error of less than \SI{0.5}{\percent}.

\section{The Proposed Controler Board and Interface}
\label{sec:implementation}
\begin{figure}[b!]
    \centering
    \vspace{-0.6cm}
    \includesvg[inkscapelatex=false, width=\linewidth, keepaspectratio]{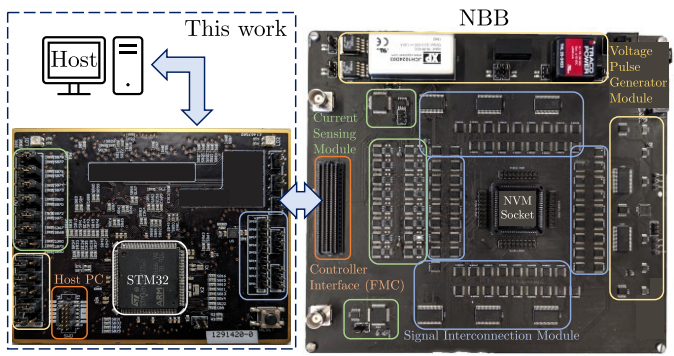}
    \caption{Integration of \ac{nbb} into automated measuring system}
    \vspace{-0.3cm}
    \label{fig:overview}
\end{figure}

The main components of the controller board developed in this paper are shown in \cref*{fig:overview}.
It can be plugged into the \ac{fmc} connector of the \ac{nbb}.
In general terms, the controller board generates the control signals for the \ac{nbb} modules.
The STM32 microcontroller manages the configurable signal interconnection module, the voltage pulse generator module, and the current sensing module.
In \cref*{fig:overview}, the control parts are highlighted in the same color as the corresponding modules of the \ac{nbb}.
All interfaces (host PC - controller board - \ac{nbb}) are marked in orange.
The controller board enables the communication between the host PC and the \ac{nvm} chip.
To be more precise, the developed board
\begin{itemize}
    \item controls the multiplexers for each operation to connect the lines to the
    \acp{bl}, \acp{wl}, and \acp{sl} of the \ac{nvm} (blue).
    \item implement \ac{rram} programming algorithms such as \ac{ispva} by
    controlling the \ac{dac} and analog switches.
    The voltage pulses must have different amplitudes, shapes, and polarities
    depending on the operation and the underlying technology (yellow).
    \item control the feedback resistors and the \ac{tia} for accurate voltage sensing (green).
\end{itemize}

\subsection*{Host PC to Control Board Communication}
For communication between the host PC and the control board,
a \ac{uart} interface  with
\ac{protobuf} for message serialization is used.
Before any command can be sent to the control board,
the setup phase of the controller must be completed to initialize the SPI channels,
the clock, and the GPIO interface.
Then the \ac{nbb} is set up, i.e.,
the IO expanders are woken up,
the \ac{dac} is cleared, the \ac{adc} converter is initialized,
the multiplexer arrays are reset by switching them to ground potential,
and the feedback resistors of the \acp{tia} are set to their highest values.

\begin{figure}[b!]
    \centering
    \includesvg[inkscapelatex=false, width=\linewidth, keepaspectratio]{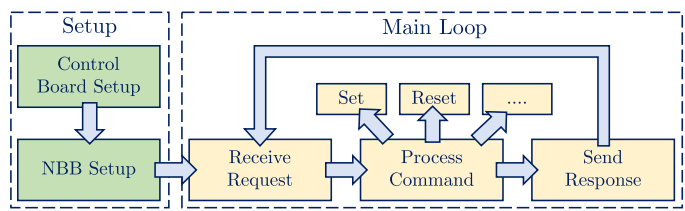}
    \caption{Setup and main loop of the control board}
    \label{fig:controlloop}
\end{figure}

After the setup phase, the main loop starts on the microcontroller
(see \cref*{fig:controlloop}).
\texttt{Receive Request} checks the host PC's request for validity and unpacks the message.
If the operation is valid, it will be executed on the \ac{nvm} chip in \texttt{Process Command}.
Finally, the result of the operation is sent back to the host PC in \texttt{Send Response}.
For example, this is the measured current of a read operation.

\subsection*{Python API}
To abstract the \ac{protobuf} format and \ac{uart} messages for the user,
we developed a host-side \ac{api}.
It is a set of Python functions that can be called
to access the \ac{nvm} cells.
A sample of the \ac{api} functions is shown in \cref*{fig:apifunctions}.
The corresponding cell can be selected with \texttt{sl} and \texttt{bl}.
Note that all \ac{api} functions have a return value which is
received by \texttt{Send Response} (for the sake of clarity, this is not shown here).
These Python functions were also used to conduct the measurements
presented in the case study in \cref*{sec:casestudy}.

\begin{figure}[b!]
\begin{python}
formCell(..., sl, bl, vGate=1.8,
    iTargetSet=80, vGateRead=1.5)
resetCell(..., sl, bl, vGateReset=2.7,
    iTargetReset=5, vGateRead=1.5)
setCell(..., sl, bl, vGate=1.5,
    iTarget=80, vGateRead=1.5)
readCell(..., sl, bl, vGateRead=1.5,
    vRead=vRead)
\end{python}
\caption{Extract of API functions}
\label{fig:apifunctions}
\end{figure}

\section{Validation of the Proposed Controller Board and Interface}
\label{sec:casestudy}

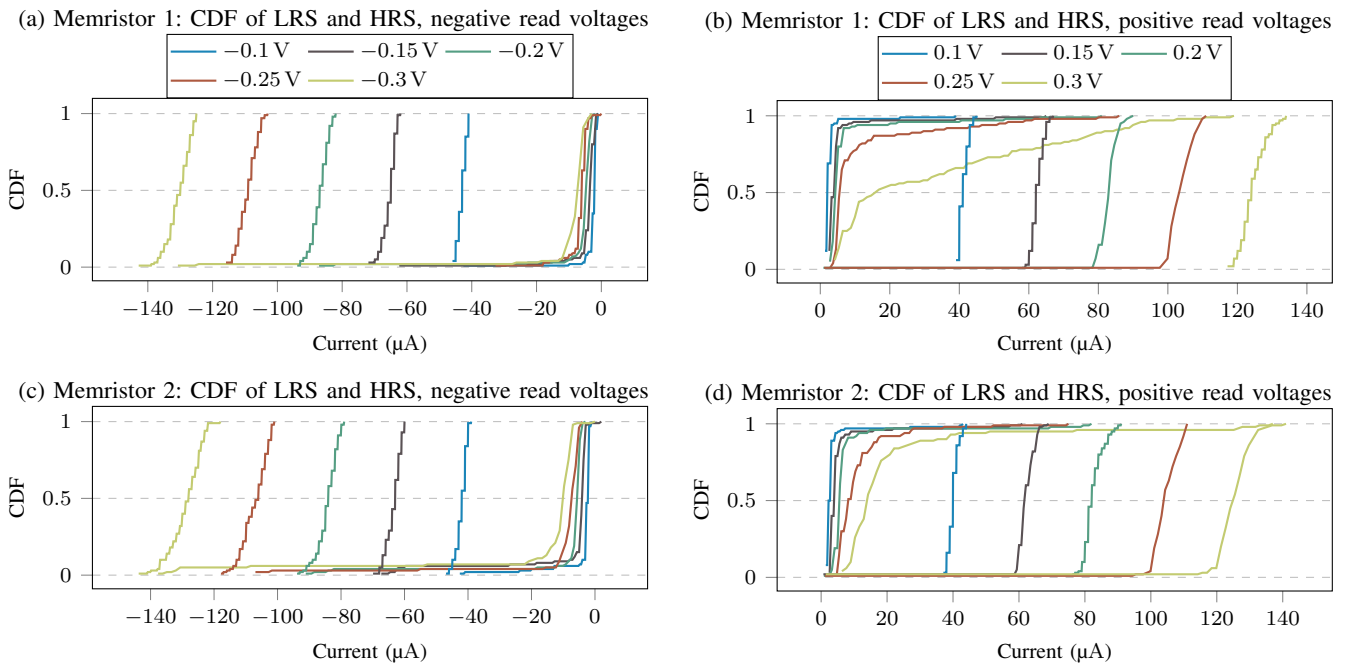
\begin{figure*}[!t]
    \centering
    \begin{subfigure}[b]{.49\linewidth}
        \centering
        \caption{Memristor 1: CDF of LRS and HRS, negative read voltages}
        \def\showlegend{1}
        \pgfplotstableread[col sep=comma]{data/chip_8/sl_1_bl_5/gen_csv/cycle-to-cycle-gen-V-0.1.csv}\curra
        \pgfplotstableread[col sep=comma]{data/chip_8/sl_1_bl_5/gen_csv/cycle-to-cycle-gen-V-0.15.csv}\currb
        \pgfplotstableread[col sep=comma]{data/chip_8/sl_1_bl_5/gen_csv/cycle-to-cycle-gen-V-0.2.csv}\currc
        \pgfplotstableread[col sep=comma]{data/chip_8/sl_1_bl_5/gen_csv/cycle-to-cycle-gen-V-0.25.csv}\currd
        \pgfplotstableread[col sep=comma]{data/chip_8/sl_1_bl_5/gen_csv/cycle-to-cycle-gen-V-0.3.csv}\curre

        \pgfdeclarelayer{bg}
\pgfsetlayers{bg,main} 
\begin{tikzpicture}
    \begin{axis}[
            General,
            xlabel={Current (\si{\micro\ampere})},
            ylabel={CDF},
            legend,
            legend style={at={(0.5,1)}, anchor=south, legend columns=3, inner sep=1pt},
        ]
        \addplot[thick, color=c0] table [x=curr_reset_x, y=curr_reset_cdf] {\curra};
        \if\showlegend1
            \addlegendentry{\SI{-0.1}{\volt}}
        \fi
        
        \addplot[thick, color=c1] table [x=curr_reset_x, y=curr_reset_cdf] {\currb};
        \if\showlegend1
            \addlegendentry{\SI{-0.15}{\volt}}
        \fi
        
        \addplot[thick, color=c2] table [x=curr_reset_x, y=curr_reset_cdf] {\currc};
        \if\showlegend1
            \addlegendentry{\SI{-0.2}{\volt}}
        \fi
        
        \addplot[thick, color=c3] table [x=curr_reset_x, y=curr_reset_cdf] {\currd};
        \if\showlegend1
            \addlegendentry{\SI{-0.25}{\volt}}
        \fi
        \addplot[thick, color=c4] table [x=curr_reset_x, y=curr_reset_cdf] {\curre};
        \if\showlegend1
            \addlegendentry{\SI{-0.3}{\volt}}
        \fi

        \addplot[thick, color=c0] table [x=curr_set_x, y=curr_set_cdf] {\curra};
        \addplot[thick, color=c1] table [x=curr_set_x, y=curr_set_cdf] {\currb};
        \addplot[thick, color=c2] table [x=curr_set_x, y=curr_set_cdf] {\currc};
        \addplot[thick, color=c3] table [x=curr_set_x, y=curr_set_cdf] {\currd};
        \addplot[thick, color=c4] table [x=curr_set_x, y=curr_set_cdf] {\curre};

    \end{axis}
\end{tikzpicture}
        \label{fig:mem1cdfneg}
    \end{subfigure}
    \begin{subfigure}[b]{.49\linewidth}
        \centering
        \caption{Memristor 1: CDF of LRS and HRS, positive read voltages}
        \def\showlegend{1}
        \pgfplotstableread[col sep=comma]{data/chip_8/sl_1_bl_5/gen_csv/cycle-to-cycle-gen-V0.1.csv}\curra
        \pgfplotstableread[col sep=comma]{data/chip_8/sl_1_bl_5/gen_csv/cycle-to-cycle-gen-V0.15.csv}\currb
        \pgfplotstableread[col sep=comma]{data/chip_8/sl_1_bl_5/gen_csv/cycle-to-cycle-gen-V0.2.csv}\currc
        \pgfplotstableread[col sep=comma]{data/chip_8/sl_1_bl_5/gen_csv/cycle-to-cycle-gen-V0.25.csv}\currd
        \pgfplotstableread[col sep=comma]{data/chip_8/sl_1_bl_5/gen_csv/cycle-to-cycle-gen-V0.3.csv}\curre
        \pgfdeclarelayer{bg}
\pgfsetlayers{bg,main} 
\begin{tikzpicture}
    \begin{axis}[
            General,
            xlabel={Current (\si{\micro\ampere})},
            ylabel={CDF},
            legend,
            legend style={at={(0.5,1)}, anchor=south, legend columns=3, inner sep=1pt},
        ]
        \addplot[thick, color=c0] table [x=curr_reset_x, y=curr_reset_cdf] {\curra};
        \if\showlegend1
            \addlegendentry{\SI{0.1}{\volt}}
        \fi
        
        \addplot[thick, color=c1] table [x=curr_reset_x, y=curr_reset_cdf] {\currb};
        \if\showlegend1
            \addlegendentry{\SI{0.15}{\volt}}
        \fi
        
        \addplot[thick, color=c2] table [x=curr_reset_x, y=curr_reset_cdf] {\currc};
        \if\showlegend1
            \addlegendentry{\SI{0.2}{\volt}}
        \fi
        
        \addplot[thick, color=c3] table [x=curr_reset_x, y=curr_reset_cdf] {\currd};
        \if\showlegend1
            \addlegendentry{\SI{0.25}{\volt}}
        \fi
        \addplot[thick, color=c4] table [x=curr_reset_x, y=curr_reset_cdf] {\curre};
        \if\showlegend1
            \addlegendentry{\SI{0.3}{\volt}}
        \fi

        \addplot[thick, color=c0] table [x=curr_set_x, y=curr_set_cdf] {\curra};
        \addplot[thick, color=c1] table [x=curr_set_x, y=curr_set_cdf] {\currb};
        \addplot[thick, color=c2] table [x=curr_set_x, y=curr_set_cdf] {\currc};
        \addplot[thick, color=c3] table [x=curr_set_x, y=curr_set_cdf] {\currd};
        \addplot[thick, color=c4] table [x=curr_set_x, y=curr_set_cdf] {\curre};

    \end{axis}
\end{tikzpicture}
        \label{fig:mem1cdfpos}
    \end{subfigure}

    \begin{subfigure}[b]{.49\linewidth}
        \centering
        \caption{Memristor 2: CDF of LRS and HRS, negative read voltages}
        \def\showlegend{0}
        \pgfplotstableread[col sep=comma]{data/chip_8/sl_5_bl_3/gen_csv/cycle-to-cycle-gen-V-0.1.csv}\curra
        \pgfplotstableread[col sep=comma]{data/chip_8/sl_5_bl_3/gen_csv/cycle-to-cycle-gen-V-0.15.csv}\currb
        \pgfplotstableread[col sep=comma]{data/chip_8/sl_5_bl_3/gen_csv/cycle-to-cycle-gen-V-0.2.csv}\currc
        \pgfplotstableread[col sep=comma]{data/chip_8/sl_5_bl_3/gen_csv/cycle-to-cycle-gen-V-0.25.csv}\currd
        \pgfplotstableread[col sep=comma]{data/chip_8/sl_5_bl_3/gen_csv/cycle-to-cycle-gen-V-0.3.csv}\curre
        \pgfdeclarelayer{bg}
\pgfsetlayers{bg,main} 
\begin{tikzpicture}
    \begin{axis}[
            General,
            xlabel={Current (\si{\micro\ampere})},
            ylabel={CDF},
            legend,
            legend style={at={(0.5,1)}, anchor=south, legend columns=3, inner sep=1pt},
        ]
        \addplot[thick, color=c0] table [x=curr_reset_x, y=curr_reset_cdf] {\curra};
        \if\showlegend1
            \addlegendentry{\SI{-0.1}{\volt}}
        \fi
        
        \addplot[thick, color=c1] table [x=curr_reset_x, y=curr_reset_cdf] {\currb};
        \if\showlegend1
            \addlegendentry{\SI{-0.15}{\volt}}
        \fi
        
        \addplot[thick, color=c2] table [x=curr_reset_x, y=curr_reset_cdf] {\currc};
        \if\showlegend1
            \addlegendentry{\SI{-0.2}{\volt}}
        \fi
        
        \addplot[thick, color=c3] table [x=curr_reset_x, y=curr_reset_cdf] {\currd};
        \if\showlegend1
            \addlegendentry{\SI{-0.25}{\volt}}
        \fi
        \addplot[thick, color=c4] table [x=curr_reset_x, y=curr_reset_cdf] {\curre};
        \if\showlegend1
            \addlegendentry{\SI{-0.3}{\volt}}
        \fi

        \addplot[thick, color=c0] table [x=curr_set_x, y=curr_set_cdf] {\curra};
        \addplot[thick, color=c1] table [x=curr_set_x, y=curr_set_cdf] {\currb};
        \addplot[thick, color=c2] table [x=curr_set_x, y=curr_set_cdf] {\currc};
        \addplot[thick, color=c3] table [x=curr_set_x, y=curr_set_cdf] {\currd};
        \addplot[thick, color=c4] table [x=curr_set_x, y=curr_set_cdf] {\curre};

    \end{axis}
\end{tikzpicture}
        \label{fig:mem2cdfneg}
    \end{subfigure}
    \begin{subfigure}[b]{.49\linewidth}
        \centering
        \caption{Memristor 2: CDF of LRS and HRS, positive read voltages}
        \def\showlegend{0}
        \pgfplotstableread[col sep=comma]{data/chip_8/sl_5_bl_3/gen_csv/cycle-to-cycle-gen-V0.1.csv}\curra
        \pgfplotstableread[col sep=comma]{data/chip_8/sl_5_bl_3/gen_csv/cycle-to-cycle-gen-V0.15.csv}\currb
        \pgfplotstableread[col sep=comma]{data/chip_8/sl_5_bl_3/gen_csv/cycle-to-cycle-gen-V0.2.csv}\currc
        \pgfplotstableread[col sep=comma]{data/chip_8/sl_5_bl_3/gen_csv/cycle-to-cycle-gen-V0.25.csv}\currd
        \pgfplotstableread[col sep=comma]{data/chip_8/sl_5_bl_3/gen_csv/cycle-to-cycle-gen-V0.3.csv}\curre
        \pgfdeclarelayer{bg}
\pgfsetlayers{bg,main} 
\begin{tikzpicture}
    \begin{axis}[
            General,
            xlabel={Current (\si{\micro\ampere})},
            ylabel={CDF},
            legend,
            legend style={at={(0.5,1)}, anchor=south, legend columns=3, inner sep=1pt},
        ]
        \addplot[thick, color=c0] table [x=curr_reset_x, y=curr_reset_cdf] {\curra};
        \if\showlegend1
            \addlegendentry{\SI{0.1}{\volt}}
        \fi
        
        \addplot[thick, color=c1] table [x=curr_reset_x, y=curr_reset_cdf] {\currb};
        \if\showlegend1
            \addlegendentry{\SI{0.15}{\volt}}
        \fi
        
        \addplot[thick, color=c2] table [x=curr_reset_x, y=curr_reset_cdf] {\currc};
        \if\showlegend1
            \addlegendentry{\SI{0.2}{\volt}}
        \fi
        
        \addplot[thick, color=c3] table [x=curr_reset_x, y=curr_reset_cdf] {\currd};
        \if\showlegend1
            \addlegendentry{\SI{0.25}{\volt}}
        \fi
        \addplot[thick, color=c4] table [x=curr_reset_x, y=curr_reset_cdf] {\curre};
        \if\showlegend1
            \addlegendentry{\SI{0.3}{\volt}}
        \fi

        \addplot[thick, color=c0] table [x=curr_set_x, y=curr_set_cdf] {\curra};
        \addplot[thick, color=c1] table [x=curr_set_x, y=curr_set_cdf] {\currb};
        \addplot[thick, color=c2] table [x=curr_set_x, y=curr_set_cdf] {\currc};
        \addplot[thick, color=c3] table [x=curr_set_x, y=curr_set_cdf] {\currd};
        \addplot[thick, color=c4] table [x=curr_set_x, y=curr_set_cdf] {\curre};

    \end{axis}
\end{tikzpicture}
        \label{fig:mem2cdfpos}
    \end{subfigure}

    \caption{\acs{cdf} of \acs{lrs} and \ac{hrs} states for different cells using negative read voltages (left) vs. positive read voltages (right).
    \label{fig:memcdf}\vspace{-0.2cm}}
\end{figure*}

To properly validate the proposed controller board and interface, we conduct a case study using a $12\times 7$ TiN/Ti/HfO\textsubscript{2}/TiN pseudo crossbar.
We evaluate the \ac{c2c} variation, read disturb, and endurance as metrics (see \cref*{sec:rrambackground}).
The crossbar is plugged into the \ac{nbb}'s socket.
The \ac{rram} devices are formed using the \ac{ispva}~\cite{perez2019toward} and the parameters provided by the vendor.
After applying a \textit{forming}-\textit{reset}-\textit{set} sequence to all cells,
the \ac{c2c} variation and the read disturb of the \ac{lrs} and \ac{hrs} are analyzed.
In the following, the \ac{ispva} algorithm is also used to program the \ac{lrs} and \ac{hrs}
(see \cref*{sec:rrambackground}).
\cref{tab:setresetparameters} lists all relevant parameters.

\begin{table}[!b]
    \centering
    \caption{\label{tab:setresetparameters}\ac{ispva} \textit{set} and \textit{reset} parameters}
    \begin{tabular}{ l | >{\raggedleft\arraybackslash}p{1.5cm}@{\,}>{\raggedright\arraybackslash}p{1cm} }
        Parameter & \multicolumn{2}{c}{Value} \\
        \hline\hline
        Top-electrode voltage \ac{lrs} ($\mathrm{V_{te}}$) & \SI{0.5}{\volt} : \num{0.1} & \si{\volt} : \SI{2}{\volt} \\
        Gate voltage \ac{lrs} ($\mathrm{V_{g,lrs}}$) & \num{1.5} & \si{\volt} \\
        Current target \ac{lrs} ($\mathrm{I_{trg,lrs}}$) & \num{80.0} & \si{\micro\ampere} \\
        \hline
        Bottom-electrode voltage \ac{hrs} ($\mathrm{V_{be}}$) & \SI{0.5}{\volt} : \num{0.1} & \si{\volt} : \SI{2}{\volt} \\
        Gate voltage \ac{hrs} ($\mathrm{V_{g,hrs}}$) & \num{2.7} & \si{\volt} \\
        Current target \ac{hrs} ($\mathrm{I_{trg,hrs}}$) & \num{5.0} & \si{\micro\ampere} \\
    \end{tabular}
\end{table}

\subsection{Cycle-To-Cycle Variation}
\label{sec:cycle2cycle}
We analyze the \ac{c2c} variation for different \textit{read} voltages, i.e., \ac{wl} voltages
(see \cref*{fig:crossbar}).
We apply negative and positive \textit{read} voltages,
specifically \SI{-0.3}{\volt}, \SI{-0.25}{\volt}, \SI{-0.2}{\volt},
\SI{-0.15}{\volt}, \SI{-0.1}{\volt}, \SI{+0.1}{\volt}, \SI{+0.15}{\volt},
\SI{+0.2}{\volt}, \SI{+0.25}{\volt}, and \SI{+0.3}{\volt}.
For each \ac{wl} voltage, we apply the \textit{set}-\textit{read}-\textit{reset}-\textit{read}
sequence \SI{100}{times} and measure the \ac{sl} current using the measurement circuitry of the \ac{nbb}
(see \cref*{sec:implementation}).

The results for two different cells are shown in \cref*{fig:memcdf}.
The \ac{cdf} of the \ac{lrs} and \ac{hrs} for the same \ac{wl} voltage share the same color.
In all measurements, both \ac{lrs} and \ac{hrs} show a higher \ac{c2c} variance
at higher \ac{wl} voltages (in terms of absolute value).
It can be seen that the measured currents in the \ac{hrs} have a smaller variance for negative \ac{wl} voltages
(\cref*{fig:mem1cdfneg,fig:mem2cdfneg}) than for positive ones (\cref*{fig:mem1cdfpos,fig:mem2cdfpos}).
The higher the (positive) \ac{wl} voltage, the more likely the \ac{hrs} swings towards \ac{lrs} when reading.
This is because the \textit{read} operation with positive voltages
has the same polarity as the \textit{set} operation (see \cref*{tab:setresetparameters}).

The same behavior has been described by the authors of~\cite{bengel2022reliability}.
They suggest reading in the \textit{reset} direction because it results in a
slower transition of the resistance state, as opposed to reading in the \textit{set} direction.
Additionally, employing lower voltages enhances the resistance stability.

\subsection{Read Disturb}
\label{sec:readstability}

\begin{figure*}[t!]
    \centering
    \begin{subfigure}[b]{\linewidth}
        \centering
        \includesvg[inkscapelatex=false, width=\linewidth, keepaspectratio]{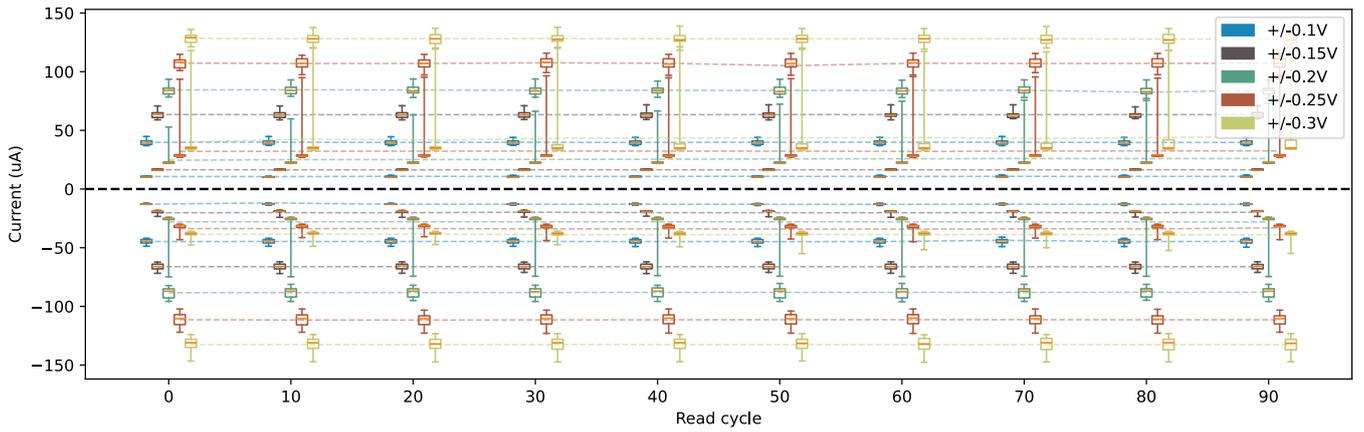}
        \caption{Read stability for cells with stable \ac{hrs}}
        \label{fig:read_stab_1}
    \end{subfigure}
    \begin{subfigure}[b]{\linewidth}
        \centering
        \includesvg[inkscapelatex=false, width=\linewidth, keepaspectratio]{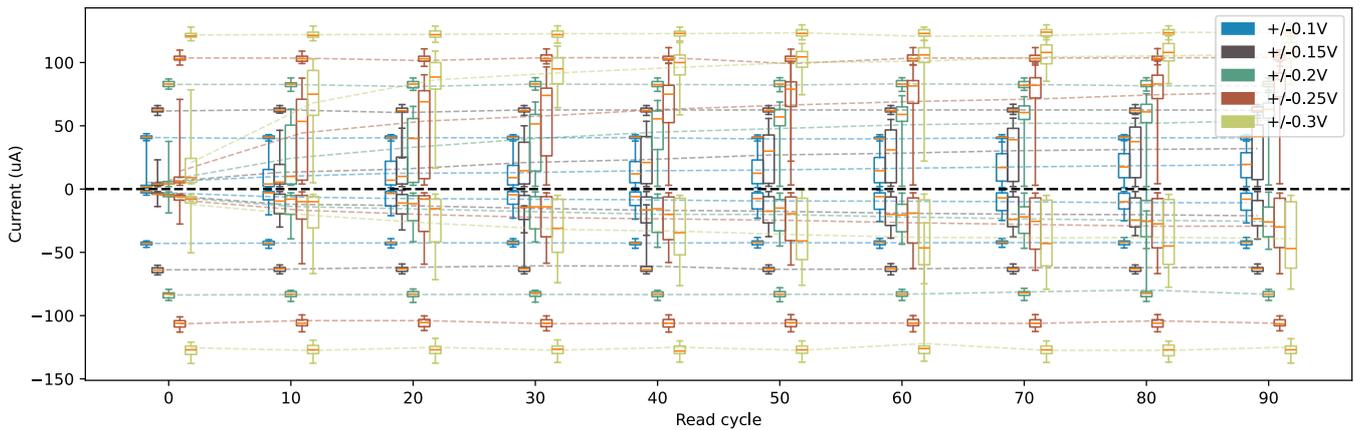}
        \caption{Read stability for cells with unstable \ac{hrs}}
        \label{fig:read_stab_2}
    \end{subfigure}
    \caption{Read stability for different \ac{rram} cells using different read voltages\label{fig:read_stab}\vspace{-0.0cm}}
\end{figure*}

Since a major application of \ac{rram} crossbars is the inference of \acp{nn},
it is important to evaluate the read disturb, which refers to the alteration of data caused by read operations.
A major advantage of \ac{cim} is that the weights, represented as \ac{lrs} and \ac{hrs},
are programmed only once and then used for multiple inferences.
Therefore, the cells must remain stable even when read several times.
We analyze the read disturb for the \ac{wl} voltages mentioned in \cref*{sec:cycle2cycle}.
For each of the \num{10} \ac{wl} voltages, we apply the sequence
\textit{set}-\textit{read(100)}-\textit{reset}-\textit{read(100)},
\SI{50}{times}, i.e, \num{1000} \textit{write} operations and \num{100000} \textit{read} operations are performed per cell.

\cref*{fig:read_stab} shows the read stability of a stable and an unstable cell.
Positive \ac{bl} currents occur as a result of positive \ac{wl} voltages and negative currents as caused by negative \ac{wl} voltages.
For clarity, only every \num{10}\textsuperscript{th} read cycle is shown.
The box plots illustrate the variation of \num{50} independent measurements.
The box boundaries represent the \SI{25}{\percent} and \SI{75}{\percent} quartiles.
The lower and upper whiskers denote the \SI{2.5}{\percent} and
\SI{97.5}{\percent} boundaries, respectively.
The median is highlighted in orange, and the dashed lines connect the mean values.
The \ac{hrs} is always the state closer to the x-axis (see also \cref*{fig:memcdf}).
We show results for cells with a stable (\cref*{fig:read_stab_1})
and unstable (\cref*{fig:read_stab_2}) \ac{hrs}.
For both cases, negative read voltages, i.e., voltages in the \textit{reset} direction, are preferred.
In cells with an unstable \ac{hrs}, the cell is continuously reprogrammed
and switches to the \ac{lrs} after just a single-digit read operation.
This effect becomes more pronounced with higher \ac{wl} voltages.
As seen in the example from \cref*{fig:read_stab_2},
\ac{hrs} and \ac{lrs} states significantly overlap with positive \ac{wl} voltages
after only \num{10} read accesses.
For read operations with negative \ac{wl} voltages, all states remain distinguishable after \num{50} accesses.
Despite separable states, the high variance in \ac{hrs} can harm computing operations.

The maximum endurance of a particular \ac{rram} technology depends on
the maximum programming energy the system can endure before failure~\cite{nail2016understanding}.
This means that each \textit{set} and \textit{reset} operation
contributes to the gradual degradation of the oxide material,
leading to failure at $N_{cmax}$.
The reported maximum endurance for \ac{oxram} technology has been documented within the range of $10^3$ to $10^9$ cycles~\cite{nail2016understanding}.
In the case of HfO\textsubscript{2}/Ti cells, oxide degradation primarily impacts the \ac{hrs},
resulting in a gradual drift of the \ac{hrs} towards the \ac{lrs}~\cite{sassine2018optimizing}.
This explains the behavior observed in \cref*{fig:read_stab_2}.
The observed $N_{cmax}$ values of the tested crossbar exhibited significant variability among different cells.
While some cells have a small $N_{cmax}$ of \num{100} measurements,
others show a stable \ac{hrs} even after \num{10000} cycles.

\section{Conclusion}
\label{sec:conclusion}
In this work, we developed a fully automated test platform
to characterize \ac{rram} cells in 1T1R crossbar arrays.
We implemented a control board that
enables communication between a host PC and the \ac{nbb}.
Using our Python-based \ac{api}, we are able to automate
test cases for cell measurements.

The proposed controller board and interface were validated in a case study.
More specifically, we analyzed the effect of read voltages on \ac{c2c} variation and read disturb in TiN/Ti/HfO\textsubscript{2}/TiN cells.
The findings reveal that reading in \textit{reset} direction achieves
better results, especially for the read disturb,
as reading in \textit{set} direction tends to swing the \ac{hrs} towards the \ac{lrs}.

Finally, this work optimizes laboratory measurement procedures,
offering considerable time savings.
Cell measurements, which may span several hours,
can now be conducted without continuous supervision.
Additionally, automated measurements ensure the consistency of data collection
by adhering to predefined and standardized procedures,
minimizing variations and human errors.
Looking ahead, the next steps are to add support for \acp{mvm}
and to automate test cases for multi-bit measurements.

\bibliographystyle{IEEEtran}
\bstctlcite{IEEEexample:BSTcontrol}
\bibliography{bibtexentry}

\begin{thebibliography}{10}
\providecommand{\url}[1]{#1}
\csname url@samestyle\endcsname
\providecommand{\newblock}{\relax}
\providecommand{\bibinfo}[2]{#2}
\providecommand{\BIBentrySTDinterwordspacing}{\spaceskip=0pt\relax}
\providecommand{\BIBentryALTinterwordstretchfactor}{4}
\providecommand{\BIBentryALTinterwordspacing}{\spaceskip=\fontdimen2\font plus
\BIBentryALTinterwordstretchfactor\fontdimen3\font minus
  \fontdimen4\font\relax}
\providecommand{\BIBforeignlanguage}[2]{{%
\expandafter\ifx\csname l@#1\endcsname\relax
\typeout{** WARNING: IEEEtran.bst: No hyphenation pattern has been}%
\typeout{** loaded for the language `#1'. Using the pattern for}%
\typeout{** the default language instead.}%
\else
\language=\csname l@#1\endcsname
\fi
#2}}
\providecommand{\BIBdecl}{\relax}
\BIBdecl

\bibitem{zou2021breaking}
X.~Zou, S.~Xu, X.~Chen, L.~Yan, and Y.~Han, ``{Breaking the von Neumann
  bottleneck: architecture-level processing-in-memory technology},''
  \emph{Science China Information Sciences}, vol.~64, no.~6, p. 160404, 2021.

\bibitem{huang2020memory}
X.~Huang, C.~Liu, Y.-G. Jiang, and P.~Zhou, ``{In-memory computing to break the
  memory wall},'' \emph{Chinese Physics B}, vol.~29, no.~7, p. 078504, 2020.

\bibitem{wei2022trends}
S.-T. Wei, B.~Gao, D.~Wu, J.-S. Tang, H.~Qian, and H.-Q. Wu, ``{Trends and
  challenges in the circuit and macro of RRAM-based computing-in-memory
  systems},'' \emph{Chip}, vol.~1, no.~1, p. 100004, 2022.

\bibitem{baroni2022energy}
A.~Baroni, A.~Glukhov, E.~P{\'e}rez, C.~Wenger, D.~Ielmini, and C.~Zambelli,
  ``{An energy-efficient in-memory computing architecture for survival data
  analysis based on resistive switching memories},'' \emph{Frontiers in
  Neuroscience}, 2022.

\bibitem{hazra2021optimization}
J.~Hazra, M.~Liehr, K.~Beckmann, M.~Abedin, S.~Rafq, and N.~Cady,
  ``{Optimization of Switching Metrics for CMOS Integrated HfO2 based RRAM
  Devices on 300 mm Wafer Platform},'' in \emph{2021 IEEE International Memory
  Workshop (IMW)}.\hskip 1em plus 0.5em minus 0.4em\relax IEEE, 2021, pp. 1--4.

\bibitem{meena2014overview}
J.~S. Meena, S.~M. Sze, U.~Chand, and T.-Y. Tseng, ``{Overview of emerging
  nonvolatile memory technologies},'' \emph{Nanoscale research letters}, 2014.

\bibitem{vetter2015opportunities}
J.~S. Vetter and S.~Mittal, ``{Opportunities for nonvolatile memory systems in
  extreme-scale high-performance computing},'' \emph{Computing in Science and
  Engineering}, 2015.

\bibitem{6724674}
N.~Raghavan \emph{et~al.}, ``{Stochastic variability of vacancy filament
  configuration in ultra-thin dielectric RRAM and its impact on OFF-state
  reliability},'' in \emph{2013 IEEE International Electron Devices Meeting},
  2013.

\bibitem{wong2012metal}
H.-S.~P. Wong \emph{et~al.}, ``{Metal--oxide RRAM},'' \emph{Proceedings of the
  IEEE}, 2012.

\bibitem{yu2011stochastic}
S.~Yu, X.~Guan, and H.-S.~P. Wong, ``{On the stochastic nature of resistive
  switching in metal oxide RRAM: Physical modeling, Monte Carlo simulation, and
  experimental characterization},'' in \emph{2011 International Electron
  Devices Meeting}.\hskip 1em plus 0.5em minus 0.4em\relax IEEE, 2011.

\bibitem{brum2021evaluating}
E.~Brum \emph{et~al.}, ``{Evaluating the Impact of Process Variation on
  RRAMs},'' in \emph{LATS}.\hskip 1em plus 0.5em minus 0.4em\relax IEEE, 2021.

\bibitem{fang2018improvement}
Y.~Fang \emph{et~al.}, ``{Improvement of HfO x-based RRAM device variation by
  inserting ALD TiN buffer layer},'' \emph{IEEE Electron Device Letters}, 2018.

\bibitem{wu2018methodology}
W.~Wu \emph{et~al.}, ``{A methodology to improve linearity of analog RRAM for
  neuromorphic computing},'' in \emph{2018 IEEE symposium on VLSI
  technology}.\hskip 1em plus 0.5em minus 0.4em\relax IEEE, 2018.

\bibitem{kempen202150x}
T.~Kempen, R.~Waser, and V.~Rana, ``{50x Endurance improvement in TaOx RRAM by
  extrinsic doping},'' in \emph{IMW}.\hskip 1em plus 0.5em minus 0.4em\relax
  IEEE, 2021.

\bibitem{liu2014uniformity}
H.~Liu \emph{et~al.}, ``{Uniformity improvement in 1T1R RRAM with gate voltage
  ramp programming},'' \emph{IEEE Electron Device Letters}, 2014.

\bibitem{baroni2021tackling}
A.~Baroni, C.~Zambelli, P.~Olivo, E.~P{\'e}rez, C.~Wenger, and D.~Ielmini,
  ``{Tackling the Low Conductance State Drift through Incremental Reset and
  Verify in RRAM arrays},'' in \emph{IIRW}.\hskip 1em plus 0.5em minus
  0.4em\relax IEEE, 2021.

\bibitem{bengel2022reliability}
C.~Bengel \emph{et~al.}, ``{Reliability aspects of binary
  vector-matrix-multiplications using ReRAM devices},'' \emph{Neuromorphic
  computing and engineering}, 2022.

\bibitem{xia2017fault}
L.~Xia, M.~Liu, X.~Ning, K.~Chakrabarty, and Y.~Wang, ``{Fault-tolerant
  training with on-line fault detection for RRAM-based neural computing
  systems},'' in \emph{DAC}, 2017.

\bibitem{xia2017stuck}
L.~Xia \emph{et~al.}, ``{Stuck-at fault tolerance in RRAM computing systems},''
  \emph{IEEE JETCAS}, 2017.

\bibitem{chen2017accelerator}
L.~Chen \emph{et~al.}, ``{Accelerator-friendly neural-network training:
  Learning variations and defects in RRAM crossbar},'' in \emph{DATE}.\hskip
  1em plus 0.5em minus 0.4em\relax IEEE, 2017.

\bibitem{berdan2015mu}
R.~Berdan, A.~Serb, A.~Khiat, A.~Regoutz, C.~Papavassiliou, and T.~Prodromakis,
  ``{A $\mu$-Controller-Based System for Interfacing Selectorless RRAM Crossbar
  Arrays},'' \emph{IEEE Transactions on Electron Devices}, 2015.

\bibitem{cayo2021design}
J.~Cayo and I.~Vourkas, ``{Design Steps towards a MCU-based Instrumentation
  System for Memristor-based Crossbar Arrays},'' in \emph{MOCAST}.\hskip 1em
  plus 0.5em minus 0.4em\relax IEEE, 2021.

\bibitem{de2020development}
R.~De~La~Fuente, I.~Vourkas, and M.~Perez, ``{On the Development of MCU-based
  ad hoc HW Interface Circuitry for Memristor Characterization},'' in
  \emph{ECCTD}.\hskip 1em plus 0.5em minus 0.4em\relax IEEE, 2020.

\bibitem{staudigl2023neuroboard}
F.~Staudigl \emph{et~al.}, ``{Work-in-Progress: A Universal Instrumentation
  Platform for Non-Volatile Memories},'' in \emph{ESWEEK}, 2023.

\bibitem{8714728}
{Guitarra, Silvana and Trojman, Lionel and Raymond, Laurent}, ``{Resistive
  Switching Model of OxRAM Devices Based on Intrinsic Electrical Parameters},''
  in \emph{2019 Latin American Electron Devices Conference (LAEDC)}, 2019.

\bibitem{poehls2021review}
L.~B. Poehls \emph{et~al.}, ``{Review of Manufacturing Process Defects and
  Their Effects on Memristive Devices},'' \emph{Journal of electronic testing},
  2021.

\bibitem{puglisi2014study}
F.~M. Puglisi, P.~Pavan, A.~Padovani, and L.~Larcher, ``{A study on HfO2 RRAM
  in HRS based on I–V and RTN analysis},'' \emph{Solid-State Electronics},
  2014.

\bibitem{grossi2016electrical}
A.~Grossi \emph{et~al.}, ``{Electrical characterization and modeling of
  pulse-based forming techniques in RRAM arrays},'' \emph{Solid-State
  Electronics}, vol. 115, pp. 17--25, 2016.

\bibitem{perez2019toward}
E.~Perez, C.~Zambelli, M.~K. Mahadevaiah, P.~Olivo, and C.~Wenger, ``{Toward
  Reliable Multi-Level Operation in RRAM Arrays: Improving Post-Algorithm
  Stability and Assessing Endurance/Data Retention},'' \emph{IEEE Journal of
  the Electron Devices Society}, 2019.

\bibitem{mao2016optimizing}
M.~Mao, Y.~Cao, S.~Yu, and C.~Chakrabarti, ``{Optimizing Latency, Energy, and
  Reliability of 1T1R ReRAM Through Cross-Layer Technique},'' \emph{IEEE
  Journal on Emerging and Selected Topics in Circuits and Systems}, 2016.

\bibitem{son2022study}
S.~Son, C.~La~Torre, A.~Kindsm{\"u}ller, V.~Rana, and S.~Menzel, ``{A Study of
  the Electroforming Process in 1T1R Memory Arrays},'' \emph{IEEE Transactions
  on Computer-Aided Design of Integrated Circuits and Systems}, 2022.

\bibitem{nail2016understanding}
C.~Nail \emph{et~al.}, ``{Understanding RRAM endurance, retention and window
  margin trade-off using experimental results and simulations},'' in \emph{2016
  IEEE International Electron Devices Meeting (IEDM)}.\hskip 1em plus 0.5em
  minus 0.4em\relax IEEE, 2016.

\bibitem{perez2021variability}
E.~Perez, M.~K. Mahadevaiah, E.~P.-B. Quesada, and C.~Wenger, ``{Variability
  and energy consumption tradeoffs in multilevel programming of RRAM arrays},''
  \emph{IEEE Transactions on Electron Devices}, vol.~68, no.~6, pp. 2693--2698,
  2021.

\bibitem{baroni2022low}
A.~Baroni \emph{et~al.}, ``{Low Conductance State Drift Characterization and
  Mitigation in Resistive Switching Memories (RRAM) for Artificial Neural
  Networks},'' \emph{IEEE Transactions on Device and Materials Reliability},
  2022.

\bibitem{sassine2018optimizing}
G.~Sassine \emph{et~al.}, ``{Optimizing Programming Energy for Improved RRAM
  Reliability for High Endurance Applications},'' in \emph{2018 IEEE
  International Memory Workshop (IMW)}.\hskip 1em plus 0.5em minus 0.4em\relax
  IEEE, 2018.

\end{thebibliography}

\end{document}